\documentclass{PoS}
\usepackage{graphicx}
\usepackage{amsmath,mathrsfs,dsfont}
\usepackage[all]{xy}
\usepackage{graphicx}
\usepackage{tensor}
\usepackage{verbatim}
\usepackage{nicefrac}
\usepackage{amsfonts}
\usepackage{amssymb,latexsym}
\usepackage{color}
\usepackage{mathrsfs,dsfont}
\usepackage{appendix}
\usepackage{slashed}
\usepackage{cite}
\usepackage{comment}
\newcommand{\Tr}{\mathrm{Tr}}
\DeclareMathOperator{\threeD}{\hspace{-2pt} {}^{(3)} \hspace{-2pt} \textit{D}}
\DeclareMathOperator{\threeR}{\hspace{1pt} {}^{(3)} \hspace{-2pt} \textit{R}}

\title{Spectral geometry approach to Horava-Lifshitz type theories: gravity and matter sectors in IR regime}

\ShortTitle{Spectral geometry approach to Horava-Lifshitz type theories}

\author{\speaker{Aleksandr Pinzul}\\

         Universidade de Bras\'{\i}lia, Instituto de F\'{\i}sica\\
         70910-900, Bras\'{\i}lia - DF, Brasil\\
         and\\
         International Center of Condensed Matter Physics\\
         C.P. 04667, Brasilia - DF, Brazil\\

        E-mail: \email{aleksandr.pinzul@gmail.com}}

\abstract{We give a brief exposition of the approach based on the methods of spectral geometry and the spectral action principle to construction and analysis of models on a foliated space-time.}

\FullConference{Proceedings of the Corfu Summer Institute 2015 "School and Workshops on Elementary Particle Physics and Gravity"\\

                 1-27 September 2015\\

                 Corfu, Greece}

\begin{document}

\section{Introduction}

Despite the significant successes of the two main candidates for the theory of quantum gravity - the superstring theory and loop quantum gravity - the final form of the theory remains elusive. This makes the development of new approaches to the problem quite desirable. For a nice review of the current status of quantum gravity, see \cite{Kiefer:2012boa}. One of such approaches has been recently proposed by Horava in \cite{Horava:2009uw}. Based on the earlier idea by Lifshitz \cite{Lifshitz}, the model appears to be at least formally renormalizable at the expense of the diffeomorphism invariance. Since the introduction, this type of models (that is naturally called Horava-Lifshitz (HL) gravity) has attracted considerable attention in the scientific community. For some recent reviews of HL gravity and some of its applications, see \cite{Padilla:2010ge,Sotiriou:2010wn,Mukohyama:2010xz}. The key ingredient of any HL-type model is a built-in foliation structure, which essentially amounts to the existence of preferred time. As a result, the fundamental symmetry of any HL theory is reduced from the full diffeomorphisms to the so-called foliation preserving diffeomorphisms (FPDiiffs). Though possessing some very attractive features of HL models, like the aforementioned renormalizability (see \cite{Barvinsky:2015kil} for recent developments), they cannot be considered as completely satisfactory due to several drawbacks, one of the most important, from our point of view, being the enormous number of free parameters in the experimentally more viable non-projectable case \cite{Blas:2010hb}. Due to this problem, a lot of unnatural fine-tuning is required to bring the model to a satisfactory agreement with the current experimental data. More problems arise if one couples matter to HL gravity in a non-minimal way.

All of the problems of any HL theory are natural continuations of its advantages: insisting that the theory is based only on FPdiffs rather on full diffeos, not only gives the hope of renormalizability, but also introduces enormous freedom in model construction.\footnote{Roughly speaking, any object respecting 3d diffeos and time reparametrization and involving up to 3 space derivatives is a legitimate building block in the action. Clearly there are many more of such objects than in a 4d diff-invariant model.} To address at least some of these problems, one needs some new principle in formulating theory, which will compensate the serious downgrading of the symmetry. We suggest that the spectral action principle \cite{Chamseddine:1996zu} could be a good candidate for this role. This approach has been extremely successful in the re-formulating the standard model coupled to general relativity (GR) in a unified geometric way \cite{Chamseddine:2008zj}. We believe that the same approach could be adopted to a much wider class of models. In particular, we advocate that it could be used to describe HL gravity naturally coupled to a Lorentz-violating matter.

In this work we review the recent efforts in this direction. More details on the calculations and discussion could be found in the original papers \cite{Pinzul:2010ct,Pinzul:2014mva,Lopes:2015bra}. The plan of the paper is as follows. We start with a very brief and schematic review of HL gravity and the spectral action principle. Then we start blending this two ingredients together: firstly, we will show how to study the physical dimension of HL-type models using the methods of spectral geometry; secondly, we discuss the notion of a geodesic motion in a space-time with a preferred foliation structure; lastly, we show how the spectral action principle can be applied to formulate in a natural way HL gravity coupled to a fermionic matter sector. We conclude with the discussion of future studies that should still be done to allow our approach to pass from a toy model to a full scale realistic theory.

\section{Original models}

In this section we will briefly describe the two main ingredients of our approach - HL gravity, i.e. a theory of gravity based on FPDiffs and the spectral action principle. We will give just enough information to facilitate the understanding of the following presentation and introduce the notations. For the detailed account on both topics see, e.g. \cite{Padilla:2010ge,Sotiriou:2010wn,Mukohyama:2010xz} and \cite{Chamseddine:1996zu}.

\subsection{Horava-Lifshitz gravity}

It is a well known fact that higher derivatives improve the ultraviolet (UV) behaviour of a theory. For example, by adding higher curvature terms to the usual Einstein action, one can make gravity perturbatevly renormalizable \cite{Stelle:1976gc}. The problem is that if one wants to keep the diffeomorphism invariance intact, one arrives at a theory which has not only higher space derivatives but also higher derivatives in time. This in its turn will lead to problems that physically are more troubling than non-renormalizability - ghosts and the loss of unitarity. So, if one wants to keep renormalizability and, at the same time, avoid the other problems, one has to consider a theory with the action that still has no more than two time derivatives and some higher order space ones. This is the basic idea in \cite{Horava:2009uw}. As an immediate consequence, the space-time acquires the structure of a foliated manifold where space and time are separated and the resulting fundamental symmetry is given by foliation preserving diffeomorphisms instead of full 4-dim diffeomorphisms. The presence of the higher order space derivatives compared to the order of the time derivative means that in the deep UV regime the theory becomes extremely non-relativistic with the anisotropic scaling for time, $t$, and space, $\vec{x}$, coordinates \cite{Horava:2009uw}
\begin{eqnarray}\label{scaling}
& &t \rightarrow a^z t  \nonumber\\
& &\vec{x} \rightarrow a\vec{x} \ .
\end{eqnarray}
Here $z$ is called the anisotropic scaling exponent. To make a theory of gravity at least formally renormalizable, $z$ should be not less than 3. The theory with $z=3$ is usually called the HL gravity.

Consider the standard ADM foliation of a space-time \cite{Arnowitt:1962hi}. The metric takes the form
\begin{eqnarray}\label{ADM}
d s^2 = (Ndt)^2 + h_{\alpha\beta} (dx^\alpha + N^\alpha dt)(dx^\beta + N^\beta dt)\ ,
\end{eqnarray}
where $N$ is the laps function, $N^\alpha$ is the shift vector, $h_{\alpha\beta}$ is the 3d metric on a leaf and $\alpha = 1,2,3$ (see the appendix in \cite{Lopes:2015bra} for the detailed discussion of 3+1 decomposition and used notations.)
Then the Einstein-Hilbert action can be written as\footnote{Due to the unsettled status of the spectral geometry and the spectral action in the pseudo-Riemannian case, see also below, we will be working with the Euclidean theory, assuming that some kind of Wick rotation is possible. Also, not to complicate things, we will be dealing with space-times without boundary. This is why there is no surface term in (\ref{EH31}).}
\begin{eqnarray}\label{EH31}
S_{EH} = \frac{M_P^2}{2} \int_M \left( K^2 - K_{\mu\nu} K^{\mu\nu} + \threeR + \Lambda_c \right) \sqrt{g} d^4 x \ .
\end{eqnarray}
Here $K_{\mu\nu} = -h_\mu^{\ \rho}\nabla_\rho n_\nu$ is the second fundamental form or the extrinsic curvature that describes the geometry of the embedding of a 3d leaf into a 4d space-time ($n_\mu$ is a 4d normal vector which has the laps function and the shift vector as the time and space components), $\threeR$ is a 3d curvature constructed out of $h_{\alpha\beta}$ in the standard way and $\Lambda_c$ is a cosmological constant. Two points are of the great importance: 1) To have 4d diff invariance all the coefficients in (\ref{EH31}) should be as they are. Changing any one of them will immediately reduce the symmetry to FPDiffs; 2) Only $K_{\mu\nu}$ contains the time derivative of the 3d metric. $\threeR$ has only space derivatives (up to the second order) of $h_{\alpha\beta}$. This shows that terms in (\ref{EH31}) with $K_{\mu\nu}$ could be considered as kinetic terms, while the rest is some sort of a potential. This leads to the following general action for Horava-Lifshitz gravity:
\begin{eqnarray}\label{HL}
S_{EH} = \frac{M_P^2}{2} \int_M \left( \lambda K^2 - K_{\mu\nu} K^{\mu\nu} + V(\threeR) \right) \sqrt{g} d^4 x \ ,
\end{eqnarray}
where $V(\threeR) = \Lambda_c + \xi \threeR + (\cdots)$ and $(\cdots)$ denotes all possible 3d invariants made of $\threeR_{\alpha\beta\gamma\delta}$ up to the cubic order (this corresponds to $z=3$, see above) as well as invariants constructed with the help of the vector $a_\alpha := \partial_\alpha N/N$. The action (\ref{HL}) represents the so-called healthy extension of the non-projectable HL gravity \cite{Blas:2010hb} (the original proposal in \cite{Horava:2009uw} did not make use of $a_\alpha$) and is the most general gravity action compatible with FPDiffs.

\subsection{Spectral geometry and spectral action}

The spectral action principle \cite{Chamseddine:1996zu} deeply roots in non-commutative or spectral geometry \cite{Connes:1994yd}. The main object in this approach to geometry is the so-called spectral triple. This construction is strongly motivated by the reformulation of Riemannian (i.e. Euclidean) geometry of a compact manifold $M$ in terms of what is called the commutative spectral triple: the commutative \textit{algebra} $C^\infty (M)$ acting on the \textit{Hilbert space} of square integrable spinors $L_2(S)$ and the standard \textit{Dirac operator}. In this case it is possible to prove the so-called reconstruction theorem \cite{Connes:2008vs} that demonstrates the complete equivalence between the standard and spectral approaches to Riemannian geometry. The non-commutative formulation, being purely algebraic, opens up the possibility for various generalizations of commutative geometry. This is achieved by generalizing in some consistent way one or all of the elements of the spectral triple (see \cite{Connes:1994yd} for some examples of the generalized geometries).

The spectral action principle can be defined for a spectral triple with some physically motivated choice of its ingredients. Schematically, this principle can be formulated as follows: the spectral triple contains not only the complete geometrical but also the complete physical information. The main problem is to choose the right spectral triple. If the choice is made, one \textit{postulates} that the dynamics of a physical system is governed by the following action
\begin{eqnarray}\label{action}
S = \Tr f \left( \frac{\mathbb{D}^2}{\Lambda^2} \right) + \langle J\psi , \mathbb{D}\psi \rangle \equiv S_{geom} + S_{matt} \ ,
\end{eqnarray}
where $\mathbb{D}$ is some (generalized) Dirac operator, $f$ is some cut-off function, $\Lambda$ is some characteristic scale and $J$ is a real structure (i.e. we should consider a \textit{real} spectral triple \cite{Connes:1995tu}). While the $S_{geom}$ part contains the description of gravity and gauge fields, the $S_{matt}$ term describes the natural and, in some sense, minimal coupling between the fermionic matter and the geometry (including gauge fields).\footnote{\label{footnote}The minimality of the matter coupling in (\ref{action}) is understood in the same way as in the usual case: $\mathbb{D}$ (with the rest of the spectral triple) completely defines the (generalized) geometry \cite{Connes:1994yd}, so coupling matter through some other Dirac-type operator would be analogous to the non-minimal (or rather non-natural) coupling.} A very important point about (\ref{action}) is that the same object, the generalized Dirac operator $\mathbb{D}$ controls both parts of the action. This is the source of the possible relations between the parameters of two seemingly unrelated sectors which we will discuss in the following sections.

One of the most impressive applications of the spectral action principle is the re-formulation of Standard Model \cite{Chamseddine:2008zj}. The spectral triple of the standard model is, in some sense, the minimal modification of the commutative spectral triple described at the beginning of this section. This modification is called the almost commutative manifold. This is a spectral triple for a geometry given by some usual (compact) Riemannian space cross some matrix geometry (i.e. the geometry, for which the algebra of functions is a matrix algebra). The appropriate (and not completely arbitrary!) choice of the matrix component leads to the completely geometric formulation of the standard model couple to general relativity. For the comprehensive exposition of this approach to the standard model, see \cite{vanSuijlekom:2015iaa}.

The main goal of this paper is to study to what extent the methods of the spectral geometry in general and the spectral action in particular can be applied to other settings, namely to theories with a preferred foliation structure.

\section{Spectral dimension \cite{Pinzul:2010ct}}

As our first application of the methods of spectral geometry to HL-type models, we would like to study the effective, i.e. as seen in an experiment, dimension of such theories. Using the definition of dimension based on a random walk, it was argued in \cite{Horava:2009if} that in deep UV regime HL gravity should behave as a 2d theory, while 4 physical dimensions will emerge only effectively in infrared (IR). This is in the complete agreement with the general analysis of \cite{Carlip:2012md} concerning the UV behaviour of a theory of quantum gravity.

To take a spectral point of view on the problem one has to understand what is a physically natural spectral triple in the presence of a foliation structure. In view of the anisotropic scaling (\ref{scaling}) it is clear that the standard Dirac operator is not natural anymore as it respects the relativistic scaling and, as a consequence, respects the full diffeomorphisms. We will discuss the general Dirac operator in the following section for the case of IR regime and for the purpose of calculating the dimension only the structure of its square, i.e. the generalized Laplacian, will suffice. It is clear that now the generalized Laplace operator will be made of all possible FPDiff invariants with the number of space derivatives going up to six, corresponding to $z=3$. So, schematically it can be written as
\begin{eqnarray}\label{completeL}
``\Delta" = \frac{\partial^2}{\partial t^2} + \Delta^3 + M_*^2\Delta^2 + M_*^4\Delta \ ,
\end{eqnarray}
where $M_*$ is some quantum gravitational scale and $\Delta$ is the usual 3d Laplacian on a leaf of the foliation. While the detailed structure of $``\Delta"$ is important for the transitional energies, we will argue that for the dimension in UV only the number of higher derivatives is relevant.

The main tool in our approach to the dimension is Weyl's theorem - one of the first results in spectral geometry

{\it Let $\Delta$ be the Laplace operator on a closed Riemannian manifold ${M}$ of dimension $\mathrm{n}$. Let $N_\Delta (\lambda)$ be the number of eigenvalues of $\Delta$, counting multiplicities, less than $\lambda$, i.e. $N_\Delta (\lambda)$ is the counting function
\begin{eqnarray}\label{counting}
N_\Delta (\lambda) := \#\{ \lambda_k (\Delta)\ :\ \lambda_k (\Delta)\leq \lambda\}\ . \nonumber
\end{eqnarray}
Then
\begin{eqnarray}\label{Weyl}
\lim_{\lambda\rightarrow\infty}\frac{N_\Delta (\lambda)}{\lambda^{\frac{\mathrm{n}}{2}}}=\frac{Vol ({M})}{(4\pi)^{\frac{\mathrm{n}}{2}}\Gamma(\frac{\mathrm{n}}{2}+1)}\ , \nonumber
\end{eqnarray}
where $Vol ({M})$ is the total volume of the manifold ${M}$.}

Weyl's theorem tells us that from the spectrum of the Laplacian we can recover not only the dimension of the manifold, but also its volume. The information about the dimension comes from the analysis of the spectrum, hence the name - spectral dimension. This definition of the dimension is very natural from the physical point of view: the spectrum of the physically relevant Laplace operator is what essentially observed in experiments. So, independently of what the mathematical (topological, Hausdorff, etc) dimension of a space-time is, its physical dimension should be associated with its spectral one. Of course, in the case of a Riemannian manifold and the standard Laplacian on it, all the dimensions will coincide. This is not the case for a non-standard choice of the Laplacian as in (\ref{completeL}). So, we would like to take a natural generalization of the Weyl's theorem for generalized Laplacians as the {\textit{definition}} of the spectral dimension.\footnote{If, due to some physical or mathematical reasons, the dimension is given or known, then the Weyl's theorem can be used to calculate the modifications of the volume due to a non-standard choice of the Laplacian. This approach was used in \cite{Gregory:2012an} to calculate the corrections due to non-commutativity to the area of a fuzzy sphere.}

\begin{figure}[htb]
\begin{center}
\leavevmode
\includegraphics[scale=0.7]{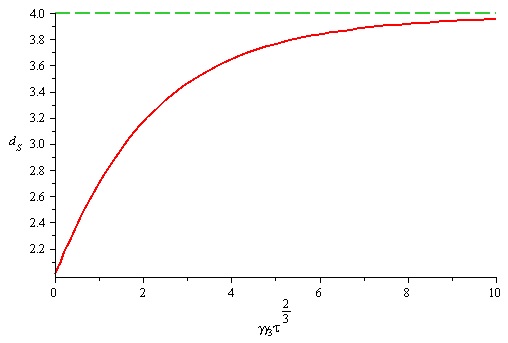}
\end{center}
\caption{The example of a smooth transition from the UV to IR regime for $z=3$ and $k=1$.}
\label{fig:scaling}
\end{figure}

First, we would like to analyze a somewhat simpler problem: the spectral dimension due to the generalized Laplacian (\ref{completeL}) where all the leaves of the foliation are flat. I.e. we consider the Laplacian \cite{Mamiya:2013wqa}
\begin{eqnarray}\label{simplifiedLap}
``\Delta" = \partial_t^2 + (-1)^{z+1}\gamma_z(\partial_i\partial_i)^z +(-1)^{k+1} \gamma(\partial_i\partial_i)^k \ . \nonumber
\end{eqnarray}
Here we take an arbitrary $z$ and one intermediate scale $\gamma$ controlling the IR behaviour, so $k<z$. The application of the Weyl's theorem essentially amounts to the calculation of the corresponding heat kernel $K(x-x';\tau )$
\begin{eqnarray}\label{heatkernel}
& &\partial_\tau K(x-x';\tau) - ``\Delta"K(x-x';\tau)=0 \\
& &K(x-x';+0)=\delta^{(4)}(x-x') \nonumber
\end{eqnarray}
and using it to calculate the spectral dimension $d_S$ according to the formula
\begin{eqnarray}
d_S = -2\frac{d\ln K(0;\tau)}{d\ln\tau}\ .\nonumber
\end{eqnarray}
The advantage of the simplification is that in this case the calculation can be performed analytically (the result in terms of Fox-Wright psi-function can be found in \cite{Mamiya:2013wqa}) giving the possibility to explicitly see the flow of the spectral dimension from $d^{UV}_S = 1 + \frac{3}{z}$ in UV to $d^{IR}_S = 1 + \frac{3}{k}$ in IR (see fig.\ref{fig:scaling}).

The complicated form of the analytic answer shows that the study of a non-flat case will be technically challenging and might require some new approach. Fortunately, if one is interested only in UV spectral dimension, one can use the approach based on the generalized $\zeta$-function \cite{Pinzul:2010ct}
\begin{eqnarray}\label{zeta1}
\zeta_{``\Delta"} (s):= \Tr (``\Delta"^{(-s)})\ . \nonumber
\end{eqnarray}
The position of the first pole of this meromorphic function is related to the so-called analytic dimension $d_a$, which is equal to the spectral one for a generalized Laplacian given by some generalized elliptic operator \cite{Higson}. The result is exactly the same as in the UV regime of the flat case:
\begin{eqnarray}\label{analytic}
d_a = 1 + \frac{D}{z} \equiv d^{UV}_S \ ,\nonumber
\end{eqnarray}
where $D$ is the number of the space coordinates and $2z$ is the order of the highest derivative. So, we see that independently of the exact form of the generalized Laplacian, the UV dimension depends on the highest spatial derivative. For the case of $D=3$, $z=3$, we get $d^{UV}_S = 2$ for \textit{any} HL model.

\section{Infrared limit}

In the previous section we saw how the spectral geometry point of view helps in defining and calculating the physical dimension of a general theory on a foliated manifold. To achieve this we used only one ingredient of a spectral triple, namely the generalized Dirac operator (actually just its square was enough). But from the definition of the spectral action (\ref{action}) it is clear that, in general, full information on the spectral triple will be essential. Here we face the following problem: while there are no conceptual complications from the side of the Hilbert space and the algebra (if one does not consider gauge fields, they will essentially remain the same as in the case of Riemannian geometry), the exact form of the generalized Dirac operator and the study of its spectral properties will change drastically. We already mentioned that even in the flat case the exact answer is very complicated, which makes it, at the moment, quite challenging to do any exact calculation for the full non-flat anisotropic case. Fortunately, from the applied point of view the study of IR regime is still very interesting. This is because infrared is defined with respect to some quantum gravitational scale and in reality includes all the observed physics and well beyond. From the point of view of the Dirac operator this means that we will have to consider the most general first order (this is infrared limit) FPDiff covariant Dirac operator. In \cite{Pinzul:2014mva,Lopes:2015bra} it was argued that such an operator can be written as
\begin{eqnarray}\label{aniso-d-1}
\mathbb{D} = \gamma^0 D_n + c_1  \threeD + c_2 \gamma^0 K + c_3 \gamma^{\alpha} a_\alpha + c_4 K + c_5 \gamma^0 \gamma^{\alpha} a_\alpha \ ,
\end{eqnarray}
where $c_i, \ i=\overline{1,5}$ are some arbitrary parameters, $K$ is the extrinsic curvature scalar and as before $a_\alpha = \frac{\partial_\alpha N}{N}$. $D_n$ is roughly a covariant derivative in the time direction and $\threeD$ is the standard Dirac operator on a 3d leaf of a foliation. The standard Dirac operator on a 4d Riemannian manifold corresponds to $c_1 = 1$, $c_2 = -1/2$, $c_3 = 1/2$ and $c_4=c_5 =0$. For all the precise definitions and notations we refer to \cite{Pinzul:2014mva,Lopes:2015bra}.

Firstly, we would like to see what is the effect of such modified geometry on the motion of a structureless point particle. Then we will address the more general question: will the spectral action based on (\ref{aniso-d-1}) lead to the infrared limit of HL gravity coupled to fermionic matter?

\subsection{Geodesic motion \cite{Pinzul:2014mva}}

Before we proceed, there are a couple of important comments we would like to make. We already commented that, at the moment, all the applications of the spectral geometry could be done only in the Euclidean regime. In particular, this is due to the complication with defining the distance for a pseudo-Riemannian geometry in the algebraic language, which can be adopted to more general geometries (this is intimately related to defining the causal structure). Actually, when calculating the spectral dimension in the previous section we already worked in the generalized Euclidean space as all the relevant operators were generalized elliptic. So, from now on we are explicitly working with the models on a foliated Euclidean space. The second comment is about the relation between geodesics and physical motion of a particle. In the Einstein general relativity with minimally coupled matter these are the same \cite{Einstein:1938yz,Dixon:1970zza,Dixon:1970zz,Hawking:1973uf}. The demonstration of this fact heavily relies on the existence of a covariantly conserved energy-momentum tensor, $T_{\mu\nu}=\frac{2}{\sqrt{-g}}\frac{\delta S_{matt}}{\delta g^{\mu\nu}}$. In HL-type theories based on FPDiff, such a conserved tensor does not exist in general \cite{Kimpton:2010xi}. So, one would like to have an alternative way to derive physical motion of a test particle starting with an action for a field coupled to gravity (which usually is more natural and fundamental than an action for a particle). While for the diff-invariant coupling we should reproduce the usual equation for geodesic motion, the situation should change in the case of the FPDiff-invariant coupling.

So, we are facing two problems: firstly, we have to understand what is a geodesic in the case of a generalized geometry; secondly, we have to find the physical motion of a test particle minimally coupled to such a geometry (see the footnote \ref{footnote}).

Let us start with the first question by showing how the geodesic distance for the usual Riemannian geometry is defined in the spectral geometry settings. As we will see, the full spectral triple will be needed to make this definition. In terms of the spectral triple data the geodesic distance between two points is shown to be equal (see, e.g. \cite{Connes:1994yd,GraciaBondia:2001tr})
\begin{eqnarray}\label{distance}
d(x,y) := \sup_{f\in {C}^\infty({M})}\left\{ |f(x)-f(y)| : \|[\mathrm{D},f]\|\leq 1\right\} \ .
\end{eqnarray}
The norm used in (\ref{distance}) is the operator norm. So, as we said before, this definition uses all the ingredients of a spectral triple: the commutative algebra $C^\infty (M)$ acting on the Hilbert space of square integrable spinors $L_2(S)$ and the standard Dirac operator $D$. When the spectral triple describes some generalized geometry, one takes (\ref{distance}) as a {\textit{definition}} of the geodesic distance.

Let us apply this to the case of our Dirac operator (\ref{aniso-d-1}). As we commented before, the algebra and the Hilbert space remain the same. So, the only component that gets modified in (\ref{distance}) is the condition $\|[\mathrm{D},f]\|\leq 1$. Using (\ref{aniso-d-1}) we can easily calculate
\begin{eqnarray}\label{Dfcomm}
[\mathbb{D},f] = \gamma^0 n^\mu \partial_\mu f + c_1 \gamma^\alpha \partial_\alpha f \ , \nonumber
\end{eqnarray}
where again $f(x)\in {C}^\infty ({M})$. As we can see, only $c_1$ coefficient contributes. But the effect due to this coefficient could be reproduced by the re-scaling of only the spatial part of the ADM metric (\ref{ADM}): $\tilde{h}_{\alpha\beta} = \frac{1}{c_1^2}{h}_{\alpha\beta}$. So we can immediately conclude that the geodesics in the generalized geometry defined by the spectral triple with the Dirac operator (\ref{aniso-d-1}) are the same as for the usual Riemannian case but with the effective metric whose spatial part was appropriately re-scaled. Now we would like to verify if the physical motion of a test particle still follows these geodesics of the generalized geometry.

As we discussed above, we would like to use a method that does not rely on the existence of a covariantly conserved energy-momentum tensor. We will achieve this by the quasi-classical analysis of the matter part of the spectral action (\ref{action})
\begin{eqnarray}\label{matter_action}
S_{matt} = \int d^4 x \sqrt{-g}\,\bar{\psi}\mathbb{D}_m \psi \ ,
\end{eqnarray}
where $\mathbb{D}_m := \mathbb{D} - \frac{mc}{\hbar}$ is the generalized Dirac operator (\ref{aniso-d-1}) with the explicitly written mass term. We already stressed that this corresponds to the minimal coupling of matter to the generalized geometry. The action (\ref{matter_action}) leads to the equation of motion of exactly the same form as in the standard case
\begin{eqnarray}\label{Diracequationgen}
\left(\mathbb{D} - \frac{mc}{\hbar}\right)\psi =0 \ .
\end{eqnarray}
The quasi-classical analysis is done in the standard way. We write $\psi$ in a standard form
\begin{eqnarray}\label{quasiclassic1}
\psi = \chi \mathrm{e}^{\frac{i}{\hbar}S} \ ,
\end{eqnarray}
where $\chi$ is a 4-spinor, while $S$ is a scalar function. Expanding $\chi$ and $S$ in $\hbar$
\begin{eqnarray}\label{quasiclassicexpansion}
\left\{
  \begin{array}{l}
    \chi = \chi_0 + \hbar\chi_1 + \cdots \ \\
    S = S_{cl} + \hbar S_1 + \cdots \\
  \end{array}
\right. \ ,
\end{eqnarray}
and plugging this into (\ref{Diracequationgen}) we arrive after some manipulations (see \cite{Pinzul:2014mva} for the details)
\begin{eqnarray}
(-n^\mu n^\nu \partial_\mu S_{cl}\partial_\nu S_{cl} + c_1^2 \gamma^\alpha \gamma^\beta \partial_\alpha S_{cl}\partial_\beta S_{cl} + m^2 c^2)\chi_0 = 0 \ . \nonumber
\end{eqnarray}
Using $\{\gamma^\alpha , \gamma^\beta\} =2h^{\alpha\beta}$ we arrive at the deformed analog of the Hamilton-Jacobi equation
\begin{eqnarray}\label{HJdefermed}
(-n^\mu n^\nu + c_1^2 h^{\mu\nu}) \partial_\mu S_{cl}\partial_\nu S_{cl} + m^2 c^2 = 0 \ .
\end{eqnarray}
We see that the whole effect of the deformation has again been reduced to the same re-scaling of the spatial metric $h^{\mu\nu}$. Formally introducing a new metric $\tilde{g}$ by
\begin{eqnarray}\label{newmetric}
\tilde{g}_{\mu\nu} = -n_\mu n_\nu + \frac{1}{c_1^2} h_{\mu\nu}  \ ,
\end{eqnarray}
we can read off the relativistic Hamiltonian of a test particle from (\ref{HJdefermed})
\begin{eqnarray}\label{Ham}
H = \tilde{g}^{\mu\nu}p_\mu p_\mu  + m^2 c^2 \ .
\end{eqnarray}
The standard analysis of a reparametriztion invariant theory with the Hamiltonian (\ref{Ham}) (see, e.g. \cite{Rovelli:2004tv}) leads to the following equations of motion\begin{eqnarray}\label{geodesicdeformed}
\frac{d^2 x^\mu}{d \tau^2}+\tilde{\Gamma}^\mu_{\nu\lambda}\frac{d x^\nu}{d \tau}\frac{d x^\lambda}{d \tau} = 0 \ ,
\end{eqnarray}
where $\tilde{\Gamma}^\mu_{\nu\lambda}$ are the Christoffel symbols calculated for the re-scaled metric $\tilde{g}_{\mu\nu}$ and $\tau$ is the modified proper time defined by $d\tau = \sqrt {- \tilde{g}_{\mu\nu}dx^\mu dx^\nu }$. This completes the proof that the physical motion of a structureless test particle follows a geodesic of the generalized geometry.

\subsection{IR Horava-Lifshitz gravity coupled to matter \cite{Lopes:2015bra}}

In this section, we would like to study the full spectral action (\ref{action}) based on the deformed Dirac operator (\ref{aniso-d-1}). The main questions that we would like to answer are 1) Does the geometric part of (\ref{action}) reproduce the IR limit of HL gravity? and 2) How does the matter part of (\ref{action}) compare to other Lorentz violating theories?

To answer the first question we need to learn how to calculate the trace that enters into the definition of $S_{geom}$ in (\ref{action}). This is done by the so-called heat kernel technique. We already used the heat kernel (\ref{heatkernel}) to calculate the spectral dimension of the flat anisotropic space-time. Now, because the deformed Dirac operator (\ref{aniso-d-1}) is not flat, we cannot expect to calculate the corresponding heat kernel exactly. Fortunately, there exists the method of finding the asymptotic expansion of a heat kernel for small proper time $\tau$ for a wide class of (generalized) elliptic operators (see, e.g. \cite{fursaev} for the detailed discussion of the heat kernel technique and its applications). Here we just summarize the main steps.

Let $P$ be some (generalized) elliptic operator of the order $m$ represented on a Hilbert space ${H}$ of square-integrable sections of some bundle over d-dimensional manifold $M$. The heat kernel of $P$ is defined by
\begin{eqnarray}\label{heat_kernel}
K(\tau,P) = \Tr_{H} e^{-\tau P} \ .\nonumber
\end{eqnarray}
It is not hard to see that this is nothing but a trace of the heat kernel defined in (\ref{heatkernel}) for $P$ instead of $``\Delta"$. This heat kernel has the well known asymptotic expansion for small $\tau$
\begin{eqnarray}
K(\tau,P) \simeq \sum_{n\geq 0} \tau^{\frac{n-d}{m}} a_n (P) \ ,\nonumber
\end{eqnarray}
where $a_n (P)$ are completely defined by some local densities, $a_n (x, P)$, known as Seeley-DeWitt coefficients:
\begin{eqnarray}\label{int_a}
a_n (P) = \int_M a_n (x, P) \sqrt{g} d^d x\ .
\end{eqnarray}
There are several techniques how to calculate these coefficients, one of the most effective being due to Gilkey \cite{Gilkey:1995mj}. Below we will give the explicit expressions for the first two coefficients for the case relevant for our purposes. The relation between the heat kernel and the trace of some function of $P$ can be established through the operator analog of the Mellin transformation. We write this relation for the case of our interest, i.e. when $d=4$ and $m=2$. The fact that we work with a manifold without boundary can be used to show that $a_n (P) = 0$ for all odd $n$ \cite{fursaev}. Then the trace of $f(P)$ is given by
\begin{eqnarray}\label{TrP}
\Tr f (P)  = \sum_{k \geq 0} f_{2k} a_{2k} (P) \ ,
\end{eqnarray}
where $f_{2k}$ are
\begin{eqnarray}
f_0  = \int\limits_0^\infty f(u) u du, \ \  f_2 =  \int\limits_0^\infty f(u) du , \ \   f_{2(k+2)} = (-1)^n f^{(n)}(0),\  k \geq 0 \ .\nonumber
\end{eqnarray}

In \cite{Chamseddine:1996zu} it was shown that choosing $P = D^2$ where $D$ is the standard Dirac operator on the twisted spinor bundle over $M$, the first three terms in (\ref{TrP}) reproduce General relativity with the cosmological constant coupled to a gauge field, which depends on the twisting.

We would like to apply this method to $P = \mathbb{D}^2$ with $\mathbb{D}$ as in (\ref{aniso-d-1}). The analysis of the previous section on the geodesic motion hints that the Dirac operator (\ref{aniso-d-1}) might look more tractable if written in terms of the re-scaled metric (\ref{newmetric}). Indeed, in this metric $\mathbb{D}$ takes form
\begin{eqnarray}\label{rescaled2}
\mathbb{D} = \tilde{\gamma}^\mu \nabla_\mu^{\tilde{\omega}} + \bigg( c_2 + \frac{1}{2} \bigg) K \gamma^0 + \bigg(c_3 - \frac{c_1}{2} \bigg) \gamma^i e_i^{\ \mu}a_\mu + c_4 K +  c_5 \gamma^0 \gamma^i e_i^{\ \mu}a_\mu  \ .
\end{eqnarray}
Here $\nabla_\mu^{\tilde{\omega}}$ is a covariant derivative on a spinor bundle with respect to the re-scaled metric and $(n^\mu , e_i^{\ \mu})$ form a set of the tetrads compatible with the ADM foliation (\ref{ADM}). It is important to notice that the first term in (\ref{rescaled2}) is the usual Dirac operator while the other terms do not contain derivatives, i.e. they are just some endomorphisms of a spinor bundle. There are well developed techniques how to deal with such operators \cite{fursaev}, so after some algebra (see \cite{Lopes:2015bra} for the details of the calculation) we arrive at the integrated Seeley-DeWitt coefficients (\ref{int_a})
\begin{eqnarray}\label{int_a_i}
a_0(\mathbb{D}^2) &=&  \frac{1}{4\pi^2} \int_M  \sqrt{g} d^4 x \ ,\nonumber\\
a_2(\mathbb{D}^2) &=& \frac{1}{48\pi^2}  \int_M \left(  c_1^2\threeR + (1-36 c_4^2) K^2 - K_{\mu\nu} K^{\mu\nu} + 12 c_5^2 a^\mu a _\mu \right) \sqrt{g} d^4 x \ .
\end{eqnarray}
Finally, using (\ref{int_a_i}) in (\ref{TrP}) we arrive at the expression for the geometric part of the spectral action $S_{geom}$
\begin{eqnarray}
S_{geom} = \frac{f_2 \Lambda^2}{48\pi^2}  \int_M \left( c_1^2 \threeR + (1-36 c_4^2) K^2 - K_{\mu\nu} K^{\mu\nu} + 12 c_5^2 a^\mu a _\mu + \frac{12 f_0 \Lambda^2}{f_2} \right) \sqrt{g} d^4 x \ .\nonumber
\end{eqnarray}
This should be compared with (the Euclidean version of) the IR limit of the action for HL gravity with the cosmological constant $\Lambda_c$ \cite{Barausse:2011pu,Barausse:2013nwa}, which is obtained by the truncation of the potential in (\ref{HL}) leaving only terms with two spatial derivatives
\begin{eqnarray}\label{IRHL}
S_{IRHL} = \frac{M_P^2}{2} \int_M \left( \xi \threeR + \lambda K^2 - K_{\mu\nu} K^{\mu\nu} + \eta a^\mu a _\mu + \Lambda_c \right) \sqrt{g} d^4 x \ .
\end{eqnarray}
It is obvious that upon the following identifications of the free parameters:
\begin{eqnarray}\label{eq:compare healthy}
M_P^2 =\frac{f_2 \Lambda^2}{24\pi^2}\ ,\ \  \sqrt{c_1} = \xi\ ,\ \  \lambda = (1-36 c_4^2)\ ,\ \ \eta = 12 c_5^2\ ,\ \ \Lambda_c = \frac{12 f_0 \Lambda^2}{f_2}
\end{eqnarray}
both actions are exactly the same. Using the map between the parameters (\ref{eq:compare healthy}) we can in principle translate the experimental bounds on $(\xi , \lambda , \eta)$ into the bounds on $(c_1 , c_4 , c_5)$ (see below).  Note that up to this order the coefficients $(c_2 , c_3)$ still do not enter. Now we pass to the consideration of the matter part of the spectral action (\ref{action}), $S_{matt}$.

The matter part of the spectral action (\ref{action}), which is given in (\ref{matter_action}), is controlled by the same Dirac operator that was used to construct its geometric part. This should be considered as a big advantage of our approach: both parts of the action contain free parameters that \textit{a priori} are independent, but now, due to the spectral action principle, we will have some relations between the parameter spaces of the geometric and the matter sectors. So we see that as was announced in the introduction, the spectral action principle does help to deal with the increased freedom due to the downgrading the full diffeomorphism symmetry to FPDiff one.

There is one more related important comment.  The matter in HL gravity has been considered before with the conclusion that, to avoid problems (fine-tuning, strong Lorentz violation, etc), one has to consider the minimal in the usual sense coupling of matter sector to the geometry, i.e. the same as in GR \cite{Pospelov:2010mp,Kimpton:2013zb}. But such a coupling is against the spirit of the spectral action principle - clearly the matter in (\ref{matter_action}) is not minimally coupled in the sense of the general relativity. But it \textit{is} minimally coupled from the point of view of the spectral action - the action (\ref{matter_action}) is really minimal and the most natural action built of some physical Dirac operator. The fine-tuning could be if not avoided completely but at least improved by the fact that the same Dirac operator controls both parts of the full action.

The matter action (\ref{matter_action}) defined with the deformed Dirac operator (\ref{aniso-d-1}) is clearly Lorentz violating. How does it fit into the general scheme for Lorentz violating theories coupled to gravity developed in \cite{Kostelecky:2003fs}? Because we have just a one-fermion toy model, the comparison will be not complete, but still it will demonstrate the main expectations from our approach. The one-fermion sector of the most general Lorentz violating theory (at the first order in the Lorentz violating parameters) is given by the following action
\begin{eqnarray}\label{SME}
S=\int d^4 x \sqrt{g} ( e^{\ \mu}_a \bar \psi \Gamma^a \nabla_\mu \psi + \bar \psi M \psi) \ ,
\end{eqnarray}
where
\begin{eqnarray}\label{LVT}
\Gamma^a &=&\gamma^a -  c_{\mu\nu} e^{ a\nu} e^{\ \mu}_b \gamma^b -  d_{\mu\nu} e^{ a\nu}e^{\ \mu}_b \gamma^5 \gamma^b - e_\mu e^{ a\mu} - i f_\mu e^{ a\mu} \gamma^5 - \frac{1}{2} g_{\lambda\mu\nu}e^{ a\nu}e^{\ \lambda}_b e^{\ \mu}_c \gamma^{bc} \ ,\nonumber \\
M &=& - m_0 + i m_5 \gamma^5 + m_\mu e_a^{\ \mu} \gamma^a +  b_\mu e^{\ \mu}_a \gamma^5 \gamma^a + \frac{1}{2}  H_{\mu\nu} e^{\ \mu}_a e^{\ \nu}_b \gamma^{a b} \ .
\end{eqnarray}
Because (\ref{SME}) is model independent, all the parameters in (\ref{LVT}) are free. It is clear that if one tries to arrive at (\ref{SME}) in the framework of some specific model, one should obtain some conditions for these parameters (see, e.g. \cite{Chatillon:2006rn} for the example of the non-commutative QED). So, our very natural expectation is that the parameters in (\ref{LVT}) will be controlled by the parameters $c_i$ from (\ref{aniso-d-1}). Indeed, comparing (\ref{matter_action}) and (\ref{SME}) one arrives at the following formulas for the non-vanishing coefficients \cite{Lopes:2015bra}
\begin{eqnarray}\label{set1}
c_{\mu \nu} = (c_1 -1) h_{\mu \nu} ,\  m_\mu = \left(\left(\frac{c_1}{2} + c_2 \right) K ,\ \left(c_3 - \frac{1}{2}\right) a_\alpha \right) ,\ m_0 = m - c_4 K ,\ H_{0 \alpha} = 2 c_5 a_\alpha \ .
\end{eqnarray}
We see that we finally get a non-trivial dependence on $c_{4,5}$. Also (\ref{set1}) shows that all the Lorentz violating parameters (except $c_{\mu\nu}$, which is somewhat trivial, see the discussion on the geodesic motion) are proportional to geometric quantities. In particular, in the limit of the flat space-time they will be zero, while in the case of strong gravity one will have strong Lorentz violation. This should have non-trivial consequences if one decides to confront our model with experiment. In the concluding section we will briefly touch on this point.

\section{Discussion and conclusions}

In this brief review, we tried to argue that the methods of spectral geometry and especially the spectral action principle may be effectively used as a guiding principle in construction of HL-type models of gravity coupled to matter. We showed that some problems, as the calculation of the spectral dimension, could be addressed in full generality, while others, at least at the moment, still can be considered only within some simplified models.

For one such toy model, a single fermion coupled to the IR Horava-Lifshitz gravity, we demonstrated the wide range of the applications of our approach based on spectral geometry. Starting with the detailed analysis of the geodesic vs physical motion of a test particle, we culminated in the \textit{derivation} of the IR limit of the HL gravity coupled to matter. One of the main advantages of our approach is the relation between the geometric and the matter parts of the spectral action. As we stressed several times, this is due to the fact that the same object, the generalized Dirac operator, is used in the construction of both terms. This could potentially have profound observational consequences. Because we are dealing with a toy model, no serious confrontation with the experimental data can be made. Still we would like to comment on what we might expect.

Typically, the parameters on the gravity side, e.g. $(\xi,\lambda,\eta)$ in (\ref{IRHL}) are much less restricted by experiment than the parameters on the matter side as in (\ref{LVT}). But now both sets of parameters are related via (\ref{eq:compare healthy}) and (\ref{set1})! So one can, in principle, use the particle physics experiment to put more restricting bounds on the gravitational parameters. E.g. in \cite{Lopes:2015bra} the bounds on the parameter $\beta:=\frac{\xi - 1}{\xi}$ are considered from the both types of experiment. While the best bound on $\beta$ coming from the gravitational experiment (based on the data from binary pulsars \cite{Yagi:2013ava,Yagi:2013qpa}) is of the order of 0.005, the bound on $\beta$ (which is related to $c_{\mu\nu}$ in (\ref{set1})) coming from the analysis of synchrotron radiation and inverse Compton scattering from astrophysical sources \cite{Altschul:2010na} is of the order of $2.5\times10^{-15}$. So, we see that if taken seriously, the spectral action principle could provide a very effective mechanism explaining different fine-tunings as well as explaining how the matter-type experiments could be used to restrict gravitational parameter space and vice versa.

Of course, there is still a long way from our toy model to a complete theory. The first and the most important step in this direction is inclusion of gauge fields. Again, the spectral action might be just the tool needed to do this in a very natural way - the gauge fields are a part of a generalized Dirac operator on some appropriately twisted spinor bundle. As we mentioned, in the case of the almost commutative geometry this leads to the geometric re-formulation of the standard model. For the case of a theory based on FPDiffs, one has to construct the twisting consistent with this symmetry.

A much more ambitious and fundamental problem is the generalization of our approach to the full theory, i.e. going beyond the IR limit. As we commented above this would require working with the most general deformation of Dirac operator by the terms up to the third order in space derivatives. This is a very challenging problem, both technically and conceptually. A lot of understanding is required and a lot of work still should be done before one could try to derive the full HL theory from the spectral action principle. But taking into account the advantages of this approach, which we tried to highlight in this work, we still hope that this will be possible to accomplish.

\end{document}